\documentclass[preprint,journal]{vgtc}       

\ifpdf
  \pdfoutput=1\relax                   
  \usepackage{graphicx}                
  \DeclareGraphicsExtensions{.pdf,.png,.jpg,.jpeg} 
\else
  \usepackage{graphicx}                
  \DeclareGraphicsExtensions{.eps}     
\fi%

\graphicspath{{figures/}{pictures/}{images/}{./}} 
\usepackage{enumitem}
\setlist{topsep=0pt, leftmargin=*}
\usepackage{microtype}                 
\PassOptionsToPackage{warn}{textcomp}  
\usepackage{textcomp}                  
\usepackage{mathptmx}                  
\usepackage{times}                     
\usepackage{cite}                      
\usepackage{tabu}                      
\usepackage{booktabs}                  

\usepackage[table,usenames,dvipsnames]{xcolor}
\definecolor{ED}{RGB}{245,201,29}
\definecolor{IC}{RGB}{25,170,178}
\definecolor{EA}{RGB}{233,114,115}

\usepackage{booktabs}
\usepackage{multirow}
\usepackage{graphicx}
\usepackage{graphicx,calc}
\newlength\myheight
\newlength\mydepth
\settototalheight\myheight{Xygp}
\newcommand*\inlinegraphics[1]{%
  \settototalheight\myheight{Xygp}%
  \settodepth\mydepth{Xygp}%
  \raisebox{-\mydepth}{\includegraphics[height=\myheight]{#1}}%
}

\usepackage{subfigure}
\usepackage{hyperref}
\DeclareUnicodeCharacter{2028}{} 
\newcommand{\name}{VizBelle}

\onlineid{1082}

\vgtccategory{Research}
\vgtcpapertype{Theoretical \& Empirical}

\title{VizBelle: A Design Space of Embellishments for Data Visualization}

\author{Qing Chen, Ziyan Liu, Chengwei Wang, Xingyu Lan, Ying Chen, Siming Chen and Nan Cao}
\authorfooter{
\item
 Qing Chen, Ziyan Liu, Chengwei Wang, Xingyu Lan, Ying Chen, and Nan Cao are with Tongji University. Emails: \{qingchen,nan.cao\}@tongji.edu.cn. Nan Cao is the corresponding author. 
\item
 Siming Chen is with Fudan University.
}

\shortauthortitle{Biv \MakeLowercase{\textit{et al.}}: Global Illumination for Fun and Profit}

\abstract{Visual embellishments, as a form of non-linguistic rhetorical figures, are used to help convey abstract concepts or attract readers' attention. Creating data visualizations with appropriate and visually pleasing embellishments is challenging since this process largely depends on the experience and the aesthetic taste of designers. To help facilitate designers in the ideation and creation process, we propose a design space, VizBelle, based on the analysis of 361 classified visualizations from online sources. VizBelle consists of four dimensions, namely, communication goal to fit user intention, object to select the target area, strategy and technique to offer potential approaches. We further provide a website to present detailed explanations and examples of various techniques. We conducted a within-subject study with 20 professional and amateur design enthusiasts to evaluate the effectiveness of our design space. Results show that our design space is illuminating and useful for designers to create data visualizations with embellishments.%
} 


\CCScatlist{ 
 \CCScat{K.6.1}{Management of Computing and Information Systems}%
{Project and People Management}{Life Cycle};
 \CCScat{K.7.m}{The Computing Profession}{Miscellaneous}{Ethics}
}

\begin{document}


\maketitle
\section{Introduction}
Visual embellishments, as a form of non-linguistic rhetorical figures, are widely used in the graphic design field, such as visual arts and advertisements \cite{BorgoEmpiricalStudyUsing2012,kang2021metamap}. 
Good visual embellishments can assist in expressing abstract notions or capture readers' attention. However, this concept has remained an inconclusive and debatable subject in the visualization domain because various researchers have different ideas and attitudes toward this issue. In Tufte's theory, visual embellishments have been defined as ``the non-data elements and redundant data elements in a graph''~\cite{TufteVisualDisplayQuantitative2001}. Following Tufte's data-ink theory, many guidelines for designing minimalist-style infographics aim to maximize the data-ink ratio by removing the the non-data part. They state that such embellishments can only complicate or even impede understanding \cite{AdornoCultureIndustrySelected2005,FewInformationVisualizationResearch2015a}. 
Meanwhile, various embellishment techniques such as shadows, textures and icons, are widely applied and presented in mass media \cite{RicheDatadrivenstorytelling2018}. Supporters argue that embellishments add value in memorization and recall to engage readers and help understand the infographics~\cite{cairo2012functional,burgio2017infographics}.

In the past few years, visualization and human-computer interaction researchers have conducted a series of studies on how embellishments influence people. Bateman et al. reported that visual embellishments could improve long-term recall without a noticeable impact on interpretation accuracy~\cite{bateman2010useful}. Borgo et al. studied visual embellishments building on the theory of conceptual structure~\cite{BorgoEmpiricalStudyUsing2012}. The results show that the effects of visual embellishments in the ``divided attention'' scenario are more prominent in facilitating users' understanding of the important concepts from visualizations. 
In this paper, we describe visual embellishment as an additional visual form for data visualization that is not directly related to visual encoding, based on prior research on the advantages of visual embellishments. Accordingly, the basic data information is still able to present to the audience when the applied form of embellishment techniques is completely stripped from the visualization. However, we could make the most of the communication power of visualization, such as quick delivery of core messages, the attitude and context of data insights, and the style that creators want to illustrate, by adding appropriate visual embellishments. 

While designing embellishments for data visualization could bring the abovementioned benefits, creating appropriate and attractive visual embellishments is a non-trivial task, especially for amateur designers who have some design experiences but have not received professional training. Most designers create their crafts directly from existing examples due to a lack of systematic guidelines for embellishment design. There is still a high demand for a useful reference to assist in the visual embellishment ideation and creation process.

To fill this gap, this paper explores the design space for various embellishment techniques so as to facilitate three major communication goals for data visualization: explore data to indicate various data insights, integrate context to enhance topic comprehension and express emotion, and enhance aesthetics to change visual styles. We first conducted expert interviews to derive key communication goals for designing visual embellishment. Then, we collected a corpus of 361 embellished visualizations from various online sources and constructed the design space by 1) coding the frequently used techniques on different objects of visualization to support the three communication goals, and 2) by summarizing them into strategies that designers could easily understand and use. We further built a website \textit{{\url{https://vizbelle.github.io/}}} to present detailed explanations and examples of the various techniques. The website presents 64 method cards derived from our design space and the corresponding existing examples from various online sources. Finally, we conducted a within-subject study with 20 professional and amateur design enthusiasts to evaluate the effectiveness of the design space and the instruction website. Qualitative and quantitative feedback was collected, and the results suggested that our design space could assist in the creation of embellishments for data visualizations. We discuss the usage patterns and future opportunities for extending our design space to further support various users' needs. 

In summary, the main contributions of this paper are as follows:
\begin{itemize}
\itemsep -1mm
    \item A design space for creating embellishments in data visualizations derived from the analysis of 361 infographics.
    \item A set of method cards derived from our corpus as teaching materials to help facilitate the design and ideation process.
    \item A workshop with 20 design enthusiasts to apply the design space to the creation of visual embellishments.
    \item Results and analysis from the qualitative and quantitative feedback of the participants from the workshop.
    
\end{itemize}

\section{Related Work}
We build on two major related areas on previous work on visual embellishment and the communications goals that help us derive the goals of embellishment creation. 

\subsection{Embellishment in Visualization}
Visual embellishment refers to the non-data elements and redundant data elements in a graph~\cite{TufteVisualDisplayQuantitative2001}. It is a controversial topic in the visualization community. Some believe that visual representations should maximize the data-ink ratio and avoid unnecessary decoration as much as possible~\cite{TufteEnvisioningInformation1990,TufteVisualDisplayQuantitative2001,AdornoCultureIndustrySelected2005,FewInformationVisualizationResearch2015a}, as they regard embellishments as chart junks that distract viewers \cite{TufteVisualDisplayQuantitative2001}, bring out visual difficulties \cite{hullman2011benefitting}, and even lead to distortion of data facts \cite{FewInformationVisualizationResearch2015a,CruzSemanticfigurativemetaphors}. This view had driven a culture of design simplification and sterilization in the data visualization community \cite{skau2015evaluation}. However, Holmes et al. \cite{holmes2000pictograms} questioned the minimalism and advocated using humour to instill affection in readers for numbers and charts. In practice, thanks to the success of the visual style of \textit{USA Today} and \textit{Time magazine}, illustrated charts and pictorial maps became popular \cite{cairo2012functional}, and had continued to retain their edge ever since \cite{burgio2017infographics}. Scholars also had come to realize that beauty does not contradict functionality \cite{NormanEmotionalDesignWhy2004}. For instance, Bateman \cite{bateman2010useful} proved that people's accuracy in describing the embellished charts was no worse than plain charts, and their recall after a two-to-three-week gap was significantly better. Few et al. \cite{FewChartjunkDebateClose2011} followed this lead, pointing out that embellishments were not always useless or damaging, and they can promote user engagement. Besides, more experiments revealed that distortion was not necessarily caused by decoration, but may be related to the lack of decluttering and focus  \cite{zacks1998reading,ajani2021declutter}. Recently, it is widely recognized by researchers that embellishments can make visualizations more memorable, more persuasive, and more engaging \cite{bateman2010useful,li2014chart,skau2015evaluation,andry2021interpreting}. Therefore, the key to resolving disputes about visual embellishments turned into using embellishments appropriately without hindering the accuracy and effiency of communicating messages \cite{KrumCoolinfographicseffective2014,LankowInfographicsPowerVisual2012}. 

Researchers in the visualization community attempted to provide theoretical bases and guidance for visual embellishment based on the overall taxonomy of abstraction and figuration \cite{burgio2017infographics,canham2010effects}. The former was helpful for exploring data insights \cite{bryan2016temporal,passonneau2014benefits,RosenbergAuToBIToolAutomatic}, and the latter could enhance comprehension and recall by embedding images as well as introducing metaphorical methods \cite{burke2009isotype,burns2021designing,Design000IconsSymbols2006,byrne2019figurative,haroz2015isotype,hou2020rhetorical,hullman2011benefitting,kang2021metamap,romat2020dear}. Yi et al. \cite{YiUnderstandingCharacterizingInsights} presented four distinctive processes of gaining insights, which inspired a lot of work dedicated to the automatic search of insights such as value, categorization, trend, and rank \cite{shi2020calliope,cui2019text,latif2021deeper}. From the figurative aspect, icons, symbols as well as pictograms with semiotic meanings are vital in reflecting social facts \cite{burgio2017infographics,burke2009isotype,engebretsen2020data} and researchers discussed various topics containing text-image relationship \cite{bateman2014text}, the vocabulary for images \cite{byrne2019figurative,hou2020rhetorical,hullman2011visualization} as well as the role they played in visualization communication \cite{avraamidou2009role,latif2021deeper,thibodeau2019role,romat2020dear}. Some noticed ``the curse of knowledge'' with regard to the differences in social consensus \cite{xiong2019curse}, thus raised higher requirements for visual embellishments.

Although previous research offered a lot of inspiration, practitioners still felt difficult to carry out findings due to the lack of systematic guidance. Moreover, their goals of designing embellishments were not explicitly classified. In this paper, we aim to fill this gap by providing users with a concise and comprehensive guide for designing visual embellishment to fulfill different communication goals.

\subsection{Communication Goals in Visualization}
Communication is a typical goal of visualization \cite{BarnardGraphicDesignCommunication2013}. Good visualizations are usually evaluated as having achieved effective communication \cite{KrumCoolinfographicseffective2014}. In traditional graphic design, which is intertwined with data visualization \cite{BertinSemiologyGraphics1983,TufteVisualDisplayQuantitative2001,TufteEnvisioningInformation1990,skau2015evaluation}, various communication goals are summarized. The most widely-accepted one is ``to be a tool for your eyes and brain to perceive what lies beyond their natural reach'' \cite{cleveland1993visualizing,FewChartjunkDebateClose2011,KosslynImageBrainResolution1996}, pointed out by Cairo \cite{cairo2012functional}. This idea helps defuse the conflict between functionality and decoration. Thus, the communication goals were enriched by works focusing on communication functions. Jacques Aumont et al. \cite{AumontImage1997} suggested three classic functions that graphic images performed and explained these functions as ``symbolic'', ``epistemic'' and ``aesthetic''. Similarly, Barnard et al. \cite{BarnardGraphicDesignCommunication2013} summarized the functions of graphic design as ``information'', ``persuasion'', ``decoration'', ``magic'', as well as ``metalinguistic and phatic''. Following previous studies, researchers in visualization continued their exploration on this topic. Krum et al. \cite{KrumCoolinfographicseffective2014} concluded the ``cool designs'' in visualization require ``engaging topic'', ``new, surprising information'', ``visually appealing and distinctive'', ``simple, focused message'',``quick and easy to read'', ``easy to share'', ``clear, easy to understand'' and ``credible data sources''. Most recently, Ajani et al. \cite{ajani2021declutter} measured the communication effectiveness of visualization on aesthetics, clarity, professionalism, and trustworthiness, focusing on epistemic and aesthetic aspects.

The goals of the cognition theory can also be a supplement to understand communication goals of data visualizations. Visualizations are often designed to improve high-level cognitive tasks \cite{figueiras2018review}, which, according to Bloom's Taxonomy \cite{best2008taxonomy}, can be described as knowledge, comprehension, application, analysis, synthesis, and evaluation. Vande et al. \cite{moere2012evaluating} suggested that information visualization is concerned with exploiting the cognitive capabilities of human visual perception in order to convey meaningful patterns and trends hidden in abstract datasets. Therefore, exploring data facts is one important communication goal for data visualizations. Bruner et al. \cite{BrunerActualmindspossible1986} unraveled the ``narrative mode'', allowing the audience to make the experience meaningful. It is especially useful in enhancing comprehension and memory \cite{avraamidou2009role}. In data visualization, the narrative mode seeks to guide the viewers and achieve narrative intents such as educating or entertaining the viewers, convincing and persuading the audience with thought-provoking opinions \cite{ojo2018patterns}. According to Slaney et al. \cite{kelliher2012tell}, data story authors could also be interested in comforting, entertain, terrorize or inform the target audience. 
Moreover, research on visualization assessment can reflect communication goals as well as validating its effectiveness \cite{saket2016beyond}. For example, Busselle and Bilandzic \cite{BusselleMeasuringNarrativeEngagement2009} created a narrative engagement measure including ``narrative understanding'', ``attentional focus'' and ``narrative presence''.

In summary, the communication goals of visualization have been enriched over time. From graphic design to information visualization, the research focus mainly lies on the inherent attributes of data and graphics with the rule of combining functionality and decoration. Cognitive science greatly expands communication goals in depth by exploiting more insights and user experiences. In the logical world, exploring distinctive facts hidden in abstract datasets can lead to various communication goals, while in the sensory world, typical narrative intents like enhancing comprehension, attracting or entertaining the viewers, and showing emotion also help to structure communication goals. Inspired by all the above work, we identify the most related communication goals for visual embellishments and further summarize them from practioners' perspective in the following expert interviews.

\section{Expert Interview}

To understand how the expert designers create visualizations with embellishments, we conducted in-depth interviews with six experienced designers. From the expert interviews, we aim to address three issues: (i) the common understanding of embellishments from practitioners and the scenarios where visual embellishments are applied to data visualizations, (ii) the potential benefits and communication goals of using embellishments, and (iii) the difficulties that designers encounter when creating visual embellishments.

\subsection{Participants} We conducted online semi-structured interviews with six experienced designers (four female), including three user experience designers (E1-E3) who have over three years' experience in designing data visualizations for BI companies, two graphic designers (E4-E5) who have six years' experience in designing visual dashboards and infographics, and one university lecturer (E6) who have over four years' experience in teaching information visualization design course. 

\subsection{Procedure} To start with, we introduced the key concepts of data embellishments and asked them to describe their expertise in the related areas. Then, we interviewed the participants with a set of prepared questions. The questions for the semi-structured interviews were informed by the analysis of the data embellishments corpus, with the goal of understanding the purpose, process, and details of creating data embellishment visualization. The example questions include, ``why and when do you use embellishment for data visualization?'', ``what factors do you consider and which methods do you usually use to design embellishments?'', ``what is the common workflow or 
``what are the difficulties you encounter when designing embellishments?'', and ``what suggestions would you give to novice designers?''. Each interview lasted for approximately 40 minutes. We recorded all the interviews via video meeting software and exported all the transcriptions. We then analyze the transcriptions and summarize the analysis results as follows.

\subsection{Result Analysis} 

\underline{\textit{Workflows and Scenarios.}}
All the experts agreed that embellishment is an essential part for visualization design in mass media. Most visualization design scenarios require embellishments, while the percentage of graphical elements greatly vary. E5 mentioned that in some data analysis scenarios for IT companies, the minimalist-style is encouraged for dashboard design where embellishments are ignored. E3 stated that ``for different target audience, the overall design style can vary a lot. Thus, the applied embellishment techniques are diverse.'' In terms of the workflow, E4 and E5 usually draw a chart first on visualization software, such as Tableau, and exported the graph to design tools, such as Adobe Illustrator or Photoshop, to add embellishments and text annotations. E1 and E2 normally start with a key concept and then explore design websites or personal databases to get inspirations, before designing the infographics to match the overall topic or to portray a specific attitude. E6 commented that the process of creating embellishments is not always linear, ``embellishment is not always considered as a fully independent component and can be added to the visualization or updated at any time.''  

\underline{\textit{Benefits and Goals.}}
From the interview summary, three major benefits and communication goals are derived. 
First, \textbf{visual embellishments can help clearly show data facts and data features}. Half the experts (E1, E2, E5) mentioned that some simple embellishments are common visual channels to express facts. For example, E1 said that ``placing prominent elements on critical data points can help readers quickly focus and display the most important information''. E2 commented that ``I usually use colors to distinguish different groups and highlight the differences between elements''. E5 recalled his experience of designing data videos and noted that ``embellishment is a critical visual channel in data charts while different types of data facts need different data embellishments, for example, if a chart shows rank, applying gradient colors is an intuitive method.''
Second, \textbf{visual embellishments can reveal the topic of the dataset or convey certain attitude}. All the experts agreed that embellishments help in enhancing understanding and memorization via using concrete images or colors. E2 said that ``to enhance comprehension and engage the audience, I often put an theme image behind or around the data content on my chart.'' Similarly, E3 mentioned that ``visual embellishments can represent certain semantic meanings which arouse emotional changes and help better understand and recall the information.'' E6 further complemented that ``some embellishments have relatively concrete information, where it forms a very good memory point. Icons and pictograms are typical embellishments as dull chart elements can thus be transformed into meaningful items.''
Third, \textbf{visual embellishments can enhance aesthetics and make the visualization more attractive}. E4 emphasized the aesthetic value of visual embellishments, ``I think embellishment is a good way to make the chart more visually appealing and interesting; the minimalist-style visualization has high data-ink ratio but no one wants to see the same style all the time''.  E4 noted that ``visual embellishments can make monotonous data visualizations more vivid, and it could attract more audience when the infographic is presented to the general public.''

\underline{\textit{Difficulties and Suggestions.}}
E1-E5 all agreed that it requires long-term practice to train novice designers. However, it would be nice to have a guidebook with abundant examples so that designers could refer to when deciding the proper embellishment style and the corresponding techniques. The experts usually browse various visualization websites and books on a regular basis, and collect well-designed examples to refer to when required. 
E6 mentioned that ``while all the forms of embellishments are applied, there are occasions where novice designers misuse those embellishment techniques.'' E6 also stated there was a lack of systematic guidance for novice designers on when and how to apply different embellishments properly.

\section{Design Space}
In this section, we present our study methodology of formalizing the design space and elaborate on the design space.

\begin{figure*}[hptb]
  \centering
  \includegraphics[width=\linewidth]{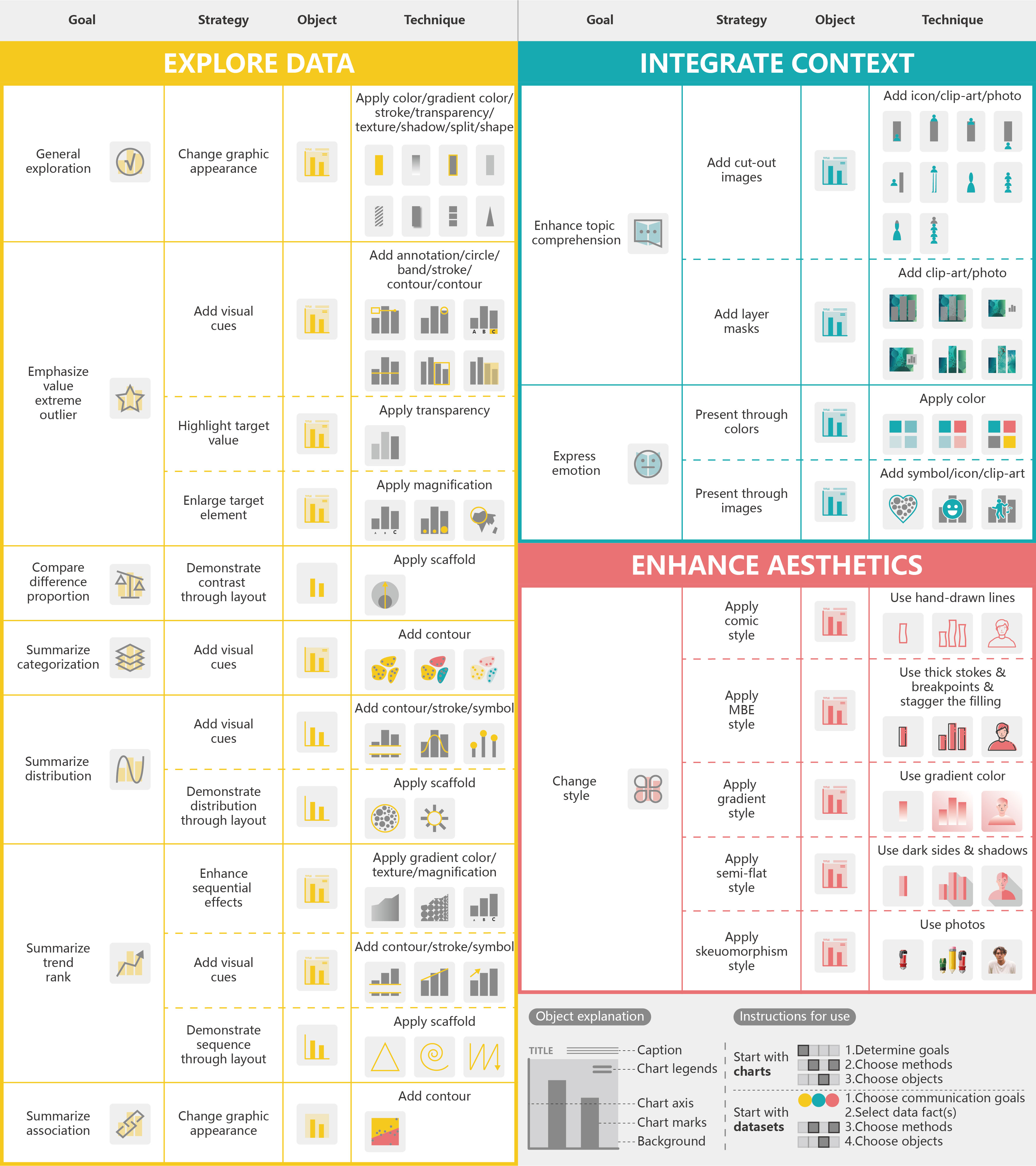}
  \caption{VizBelle design space for creating visual embellishments for data visualization.}
  \label{fig:designspace}
\end{figure*}

\subsection{Methodology}
\subsubsection{Data Collection and Selection}
We first attempted to collect infographics from previous papers~\cite{segel2010narrative,ren2017chartaccent}. However, some resource links are partially missing or unavailable. We thus surveyed up-to-date infographics from 18 sources, including established news agencies (e.g., The Guardian, Economics, FiveThirtyEight, Reuters, The New York Times, and BBC News) and design websites (e.g., Behance, Flowingdata, DailyInfographic).

To gain a deeper understanding of how visual embellishments achieve communication goals, we exclude those minimalist-style infographics and collected 361 embellished work from online sources. Two of the authors first screened all the samples independently to ensure that all the selected samples (1) contained specific data facts, (2) effectively achieved communication goal, and (3) applied one or multiple explicit embellishment techniques (that is, additional decorations or transformations other than the necessary visual encoding). Then, the two authors met for multiple sessions to compare their screening results, and discuss mismatches until achieving consensus. 
The complete list and all the samples can be found on our website \textit{{\url{https://vizbelle.github.io/}}}. Next, we screened and classified the corpus according to the communication goals. 
Based on the related work surveyed and the result analysis from the expert interviews, we summarize the communication goals of visual embellishments into three major categories: a) exploring data, b) integrating context, and c) enhancing aesthetics. Among them, exploring data focuses on the means of adding auxiliary lines or changing color to the important information on the visual display target, so as to highlight the information, direct user attention to the key information, and help users better understand the data encoding. Integrating context involves arousing the users' awareness and association with the target theme, by adding topic-related materials to emphasize the context, or by mobilizing user emotions through the introduction of relevant icons, clip-arts or photos. Enhancing aesthetic aims to unify the graphic elements into a set of visual styles, to promote the visual characteristics and impact, in order to attract target audience on suitable occasions. 
We found that when the content of an infographic involves more than one visualization, its communication goals vary. The goal of the infographic focuses more on the correlation between the two sets of data, such as hierarchy and causality, which is usually achieved by layout techniques, such as alignment and juxtaposition. Therefore, we filtered out the infographics with such multi-chart content.

\subsubsection{Coding Scheme}
At first, we coded the corpus from three aspects: data facts, communication goals, and embellishment techniques. After coding, we found out that various data facts and different communication goals can have a good correspondence relationship. We then further divided the communication goals into more specific categories. However, embellishment techniques tended to form a system of their own and ended up being mere aesthetic tools. Accordingly, we adjusted the coding relationships to reflect the implementation of communication goals through a summary of embellishment strategies. We also discussed with several expert designers from previous interviews in the iterative process of framework design. The general idea is to find the commonality of the various sets of embellishment techniques, and to redefine it with the corresponding communication goals, speculating why the designers conducted such actions to achieve their design purposes. After several coding iterations, we finally established four dimensions to describe the creation of embellishment: communication goal, strategy, object, and technique, which will be explained in detail in the following sections. 
There are multiple sub-categories in each major communication goal. The same technique can be applied to serve different communication goals. The low-level embellishment techniques are grouped in each into high-level strategies for each specific communication goal.

\subsection{Design Space Overview}
In this section, we describe the definitions and functions of each dimension in our design space before diving into the individual items in each communication stage. Notably, the strategy, object, and technique dimensions center around the goal dimension to provide approaches to achieve the corresponding communication goals.

\subsubsection{Goal}
Under the guidance of the communication goals, we explain how to select and organize various elements in charts (i.e., data visualizations with statistic meanings). It can be formally defined as methods for selecting and organizing data facts as well as context and aesthetic materials to accomplish certain communication goals. The first major goal is to explore data, unless otherwise noted, the explored data facts can be of any type (e.g., trend, value, difference, etc.) \cite{shi2020calliope,wang2019datashot} individually, and can also appear in combinations, as observed from our corpus. In terms of the ``integrating context'' goal, we identify two specific communication goals. One goal describes how to enhance topic comprehension of the chart, the other one focuses on expressing emotions\cite{cruz2015wrongfully}. Finally, we present the concept of modifying visual styles for the ``enhancing aesthetics'' goal inspired by Lau's theory \cite{lau2007towards}. In practice, visualization designers usually have one or multiple communication goals in mind and require decisions on appropriate approaches to achieve their goals.

\subsubsection{Strategy}
The strategy dimension is used to summarize and classify techniques to help users improve their understanding. Strategy describes the logic of applying certain techniques to meet certain goals. We propose a set of strategy types (e.g., add visual cues, demonstrate through layout, present through colors, and apply comic style, etc.) that are delineated by their effects and characteristics in infographics. The visual cues that we mentioned here are elements that are added on top of the model to modify its graphical appearances \cite{cruz2015wrongfully}. The idea of demonstrating through layout comes from the concept of visual information flows (VIF) from \cite{lu2020exploring}. This concept refers to the underlying semantic structure that links the
graphical data elements to convey the information and story
to the user. Finally, we identified 21 strategies by analyzing the 64 techniques used in our corpus. 

\subsubsection{Technique}
The technique dimension addresses how goals are presented to the audience. These machanisms can be detailed as low-level visual design techniques that improve visualization presentation. This dimension covers a wide range of visual design strategies such as adding visual cues, adding images, and changing the styles used in the charts. We use icons to represent the design techniques to avoid using long text narrations. Some techniques are more suitable for computer-aided creation, such as presenting a cutout image for the goal concrete character. Meanwhile, other techniques contain graphic design ideas, such as layout scaffold to demonstrate the difference.

\subsubsection{Object}
To better describe the target object for applying embellishment techniques, we refer to the basic components of visualization charts according to the visualization taxonomy. In Borkin's definition of visualization taxonomy \cite{borkin2015beyond}, the visualization elements can be classified to data encoding, data-related components (e.g., axes, annotations, legends, etc.), textual elements (e.g., title, axis labels, paragraphs, etc.), human recognizable objects (HRO), and graphical elements with no data encoding function. The roles of visualization taxonomy as ``visual elements'' and ``component types'' can also be viewed as objects in data visualization. We thus organize and identify six basic object types. The objects refer to chart components that the users can apply the embellishment techniques to, as are shown in the bottom right corner in Fig. \ref{fig:designspace}, including title, caption, chart legends, chart axis, chart marks, and background. 

\subsection{Explore Data}
We identified seven specific goals that can be applied at the data exploration stage as shown in the left column in Fig. \ref{fig:designspace}.

\textbf{\textcolor{ED}{General exploration.}} In the process of technique classification, we found that a couple of techniques can be applied to any type of data fact. We thus merged these universal techniques into one category to serve the general exploration purpose. Most are basic techniques so that users can easily apply them to the corresponding visual object. 

\textit{Strategy}. All the techniques in this category can be considered as changing the graphic appearance of the original object. 
\textit{Object}. All the objects can be applied to. 

\textit{Technique}. The techniques aim for general exploration are straightforward and widely used in practice. These techniques include changing color (solid), applying gradient color, apply a stroke, applying transparency, applying texture, applying shadow, applying split, and transforming shape. Multiple techniques can often be applied together to emphasize the key information.

\textbf{\textcolor{ED}{Emphasize value/extreme/outlier.}} 
These three data facts, value, extreme, and outlier, are frequently exposed as the focus attention of a visualization. Therefore, this goal is to strengthen this effect by applying three major embellishment strategies. 

\textit{Strategy}. Appropriate strategies include adding visual cues, highlighting target value, and enlarging target elements, all of which guide viewers' attention to certain areas. 

\textit{Object}. All the objects can be applied to. 

\textit{Technique}. Users can add annotation, add a circle, overlay a band, add a stroke, overlay contours to the target data, or apply transparency and magnification when exhibiting the emphasized value/extreme/outlier.

\textbf{\textcolor{ED}{Compare difference/proportion.}}
Visualizations that aim at comparing difference/proportion always display data facts that differ significantly. As a result, they reveal contrasts between two aspects of the dataset, building conflicts to impress viewers. 

\textit{Strategy}. Demonstrating two different values under the same measurement following certain layout rules can further present their difference or proportion.

\textit{Object}. All the objects can be applied to. 

\textit{Technique}. We discover that by superimposing data in scaffolds such as bottom alignment or center alignment on top of the picture, a clear comparison is shown.

\textbf{\textcolor{ED}{Summarize categorization.}} This goal poses the classification results directly to the audience to enable quick understanding. 

\textit{Strategy}. The corresponding strategy is adding visual cues to show various categories more clearly. 

\textit{Object}. All the objects can be applied to. 

\textit{Technique}. We identified three embellishment techniques containing adding contours with different shapes and colors which reflect categorization.

\textbf{\textcolor{ED}{Summarize distribution.}} This goal focuses on providing the distribution information of the data. 

\textit{Strategy}. In information visualization, distribution can be highlighted by using appropriate visual cues or being arranged within a certain shape or along a scaffold. 

\textit{Object}. It can be used on chart marks and chart axis. 

\textit{Technique}. The commonly identified technique approaches to convey the distribution information are adding contours, strokes, or symbols. Moreover, chart elements can also be shown by being arranged around a basic shape.

\textbf{\textcolor{ED}{Summarize trend/rank.}} Summarizing trend/rank requires sorting things according to certain criteria and revealing them one by one to establish a sense of progression or regression. 

\textit{Strategy}. The corresponding strategy typically includes enhancing sequential effects, adding visual cues, and demonstrating sequence through the layout.

\textit{Object}. The first two strategies can be applied to all the objects, while the last one is more limited to chart marks and chart axes. 

\textit{Technique}. We observed nine techniques that can enhance feelings of surprise when presented with trend or rank. Applying gradient color, texture, magnification, adding contours, strokes, and symbols are applicable ways that are easy to understand. Applying scaffold means the audience can layout the chart marks around the selected shapes.

\textbf{\textcolor{ED}{Summarize association.}} Summarizing association means showing the positive correlation or negative correlation of data. 

\textit{Strategy}. The corresponding strategy is changing graphic appearance.

\textit{Object}. It can be applied to chart marks and background. 

\textit{Technique}. The technique of adding contours can be used to highlight or fill both sides of the trend line.

\subsection{Integrate Context.}
We identified two specific goals that can be applied at the context integration stage as shown in Fig. \ref{fig:designspace}).

\textbf{\textcolor{IC}{Enhance topic comprehension.}} This goal focuses on providing contextual information of the data, such as the motivation or the background of the story. 

\textit{Strategy}. In journal graphics, topic information often refers to the time, place, characters, or environments of the events. We discovered that adding cut-out images and layer masks are the most commonly used strategies. 

\textit{Object}. The corresponding techniques can be used on any object. 

\textit{Technique}. The use of related icons, clip-arts, and images is a generally acknowledged way of conveying abstract information. Furthermore, when the topic is related to data, it is more likely to be included as a part element; while, when the topic is tied to a dataset, it is more likely to be added as a full element.
 
\textbf{\textcolor{IC}{Express emotion.}} To express emotion needs to build empathy \cite{DuarteDatastoryexplain2019}, by arousing the audience's feelings through the embellishment designs. 
\textit{Strategy}. Its associated techniques seek to make a powerful impact on the audience, allowing them to sense the creator's emotions through colors and pictures.

\textit{Object}. All the objects can be applied to. 

\textit{Technique}. These techniques can reduce the cognitive burdens of the audience, enhance comprehension, and boost their emotional involvements by applying color and adding symbols, icons, or clip-arts.

\subsection{Enhance Aesthetics.}
We identified one specific goal that can be applied at the aesthetics enhancement stage as shown in Fig. \ref{fig:designspace}. 

\textbf{\textcolor{EA}{Change style.}} This goal aims at improving attractiveness at first glance and matching the appropriate application scenarios. 

\textit{Strategy}. It requires designers to use the external domain knowledge to select a proper style. The strategies can vary from case to case. In this design space, we summarize five commonly used strategies from mainstream design websites: comic style, MBE style, gradient style, semi-flat style, and skeuomorphism style. 

\textit{Object}. All the objects can be applied to. It is ideal to keep the overall design style consistent. 

\textit{Technique}. We present the techniques by providing examples and key elements of each style.


\section{Workshop}
To evaluate the design space, we organized a workshop with 20 design enthusiasts to assess: G1) usefulness of design space, G2) easy to use, and G3) user satisfaction of final outcomes. We conducted the within-subjects study to reduce the possible effect of individual differences in the design process. The baseline is the outcome when the participants designed visual embellishments without the proposed design space and method cards.

\subsection{Hypotheses}
We make the following four hypotheses to investigate both the experience in using the design space and the outcome satisfaction. 

H1: The embellishment techniques adopted by users after using the design space are richer and more diverse than before (G1). 

H2: After using the design space, user satisfaction of the design outcomes and their performance will increase (G3). 

H3: Professionals can understand our design space quickly and accurately, and amateurs can achieve similar results after training (G2). 

H4: After using the design space, the outcome improvement of amateurs will be significantly greater than that of experts (G1). 

\subsection{Participants}
We recruited 20 participants (14 females) aged between 20 and 30 by disseminating advertisements on online social media platforms. All the participants were university students or have obtained the Bachelor degree. Before the workshop, half of the participants were amateur designers who had studied data visualization in classes or at work but without practical experience; the other half were familiar with graphics design since their professions are closely related to design and data visualization. In both the amateur and the expert group, their professions and majors can be broadly categorized into design (including industrial design, architecture and urban planning, journalism) and technical backgrounds (including computer science, chemical engineering, and data science). This classification is later used to analyze usage patterns of the design space by users with different backgrounds. Each participant was asked to sign an ethics approval before the workshop.

\subsection{Teaching Materials and Datasets}
During the workshop, we provided a set of method cards as the teaching materials of our design space. The Napa Cards \cite{NAPACardsNarrative} and IDEO Method Cards \cite{MethodCards} were used as inspiration for the method cards' design. The purpose was to provide examples to help participants comprehend each category of our design space and to inspire them to employ goals, methods, objects, and techniques in data embellishment creation. Each method card depicts a strategy or technique and a variety of related information from three perspectives: how to apply it, why it is helpful, and instances of its application in various objects.
Our website \footnote{\url{https://vizbelle.github.io/}} presents all the method cards, the example corpus we collected, and four datasets covering different topics: the Covid-19 India statewise dataset, the 2021 Tokyo Olympics dataset, the Netflix original films \& IMDB score dataset, and the food nutrient dataset. We expected that the wide range of data facts extracted from the four datasets could result in a variety of visual representations. In addition to the raw data, we provided various types of extracted data facts (e.g., value, outlier, rank, distribution, etc.  \cite{shi2020calliope,wang2019datashot}) and their corresponding visualizations to facilitate participants' design process of visual embellishments based on the suggestions from previous studies \cite{kandogan2012just,ren2017chartaccent}.

\begin{figure*}[tb]
  \centering
    \subfigure[]{
    \includegraphics[height = 5cm]{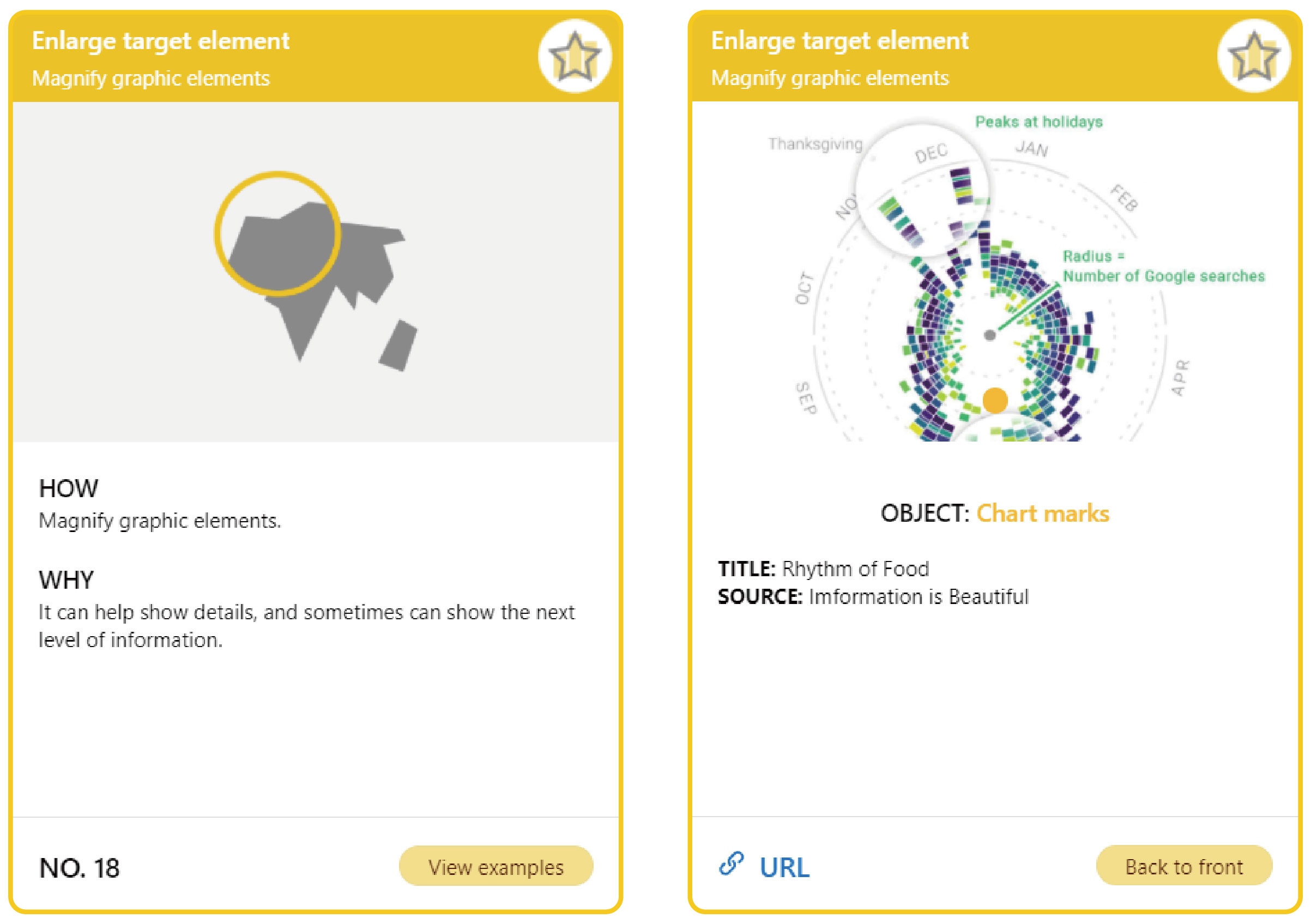}}\hspace{0.5cm}\subfigure[]{
    \includegraphics[height = 5cm]{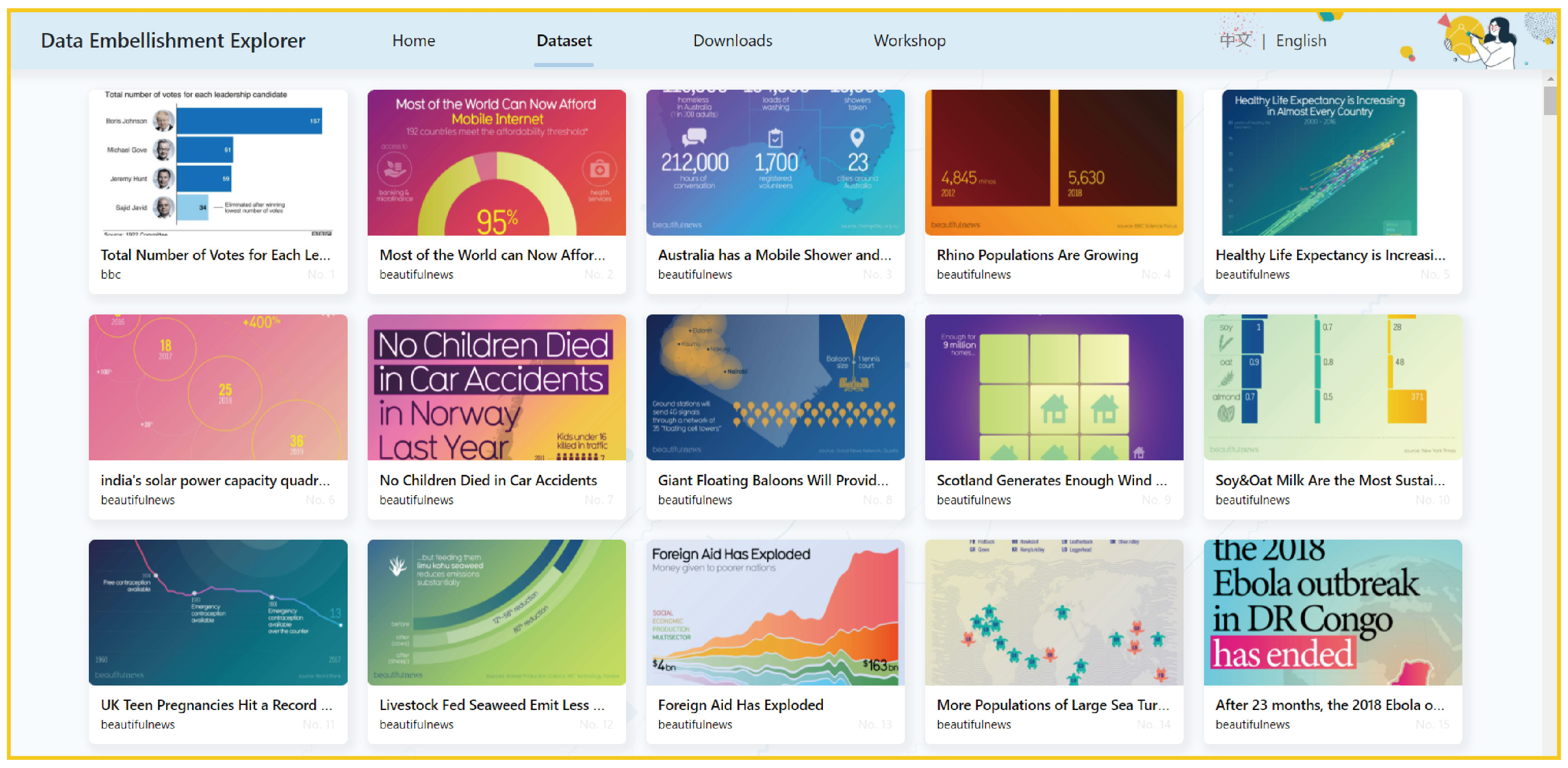}}
  \caption{The website interface of \name: (a) shows an example method card for explaining a technique in the design space: front and back, (b) presents the datasets of selected infographics from online sources.}
  \vspace{-2mm}
  \label{fig:chart}
  
\end{figure*}

\subsection{Procedure}
\begin{figure*}[tb]
  \centering
  \includegraphics[width=0.9\linewidth]{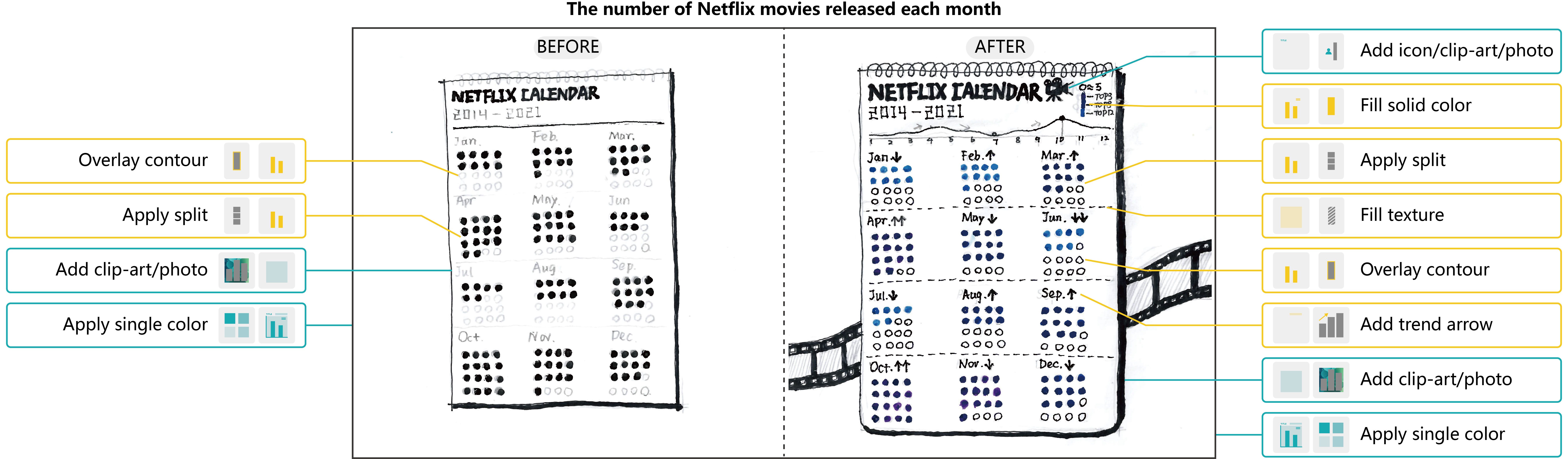}
  \includegraphics[width=0.9\linewidth]{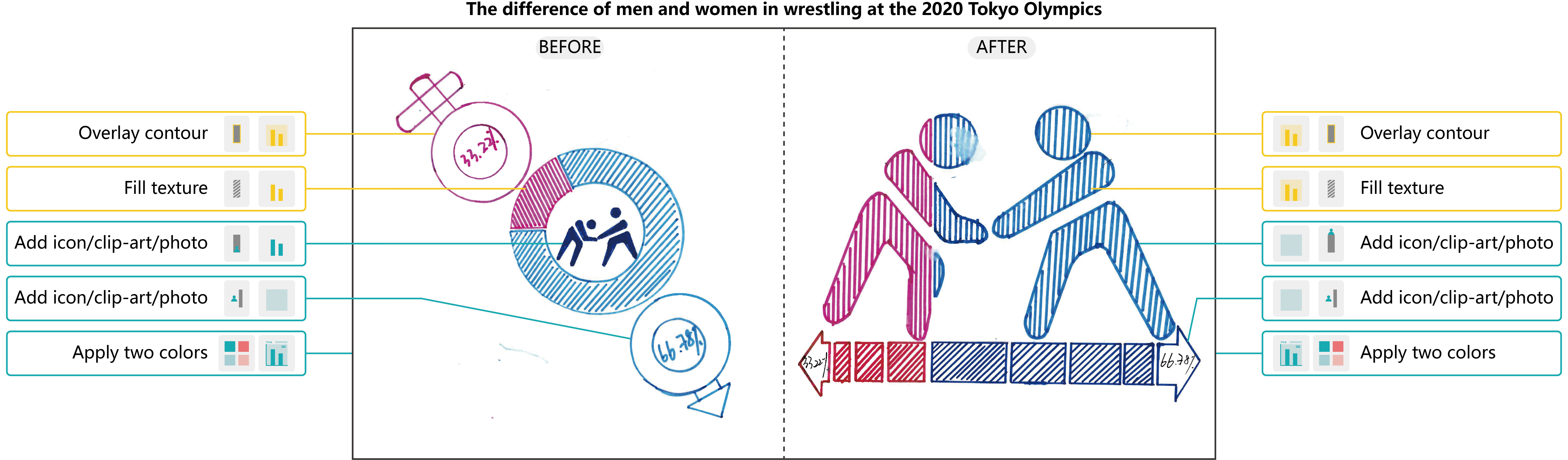}
  \caption{Design outcomes from the workshop (before and after the introduction of VizBelle: \name: (a) the number of Netflix movies released each month (designed by amateur participants), (b) The difference of men and women in wrestling at the 2020 Tokyo Olympics (designed by expert participants). The corresponding communication goals, objects and techniques are noted beside each chart.}
  \label{fig:workshop}
  \vspace{-4mm}
\end{figure*}

The workshop lasted approximately four hours. For constructing the chart, we started with a 30-minute introduction to data visualization, data embellishment, and the concept of data fact. We exhibited two example graphics that were produced using data embellishment to show the audience how data embellishment could be applied to graphics. In the communications, we split the charts into different object parts and provided a detailed account of how each section was constructed. Following the introduction, all the participants were randomly divided into ten groups. Each group was assigned the task of creating graphics with data adornment using either the provided datasets or any internet sources they choose. 
We gave each group 60 minutes to explore the dataset and create two visualizations with key points at each stage before diving into the design space to conduct a thorough investigation into how our design space could help designers construct visual charts from goal planning to final accomplishment. Following that, we demonstrated how to use our design space as well as the technique cards on our website. The participants were then given free rein to enhance their graphics in our design area. We told each group to sketch their graphics on the template to showcase the visualizations of their charts after they were finished (as shown in Fig. \ref{fig:workshop}). Participants were also given the freedom to review down any visual design ideas they had that couldn't be demonstrated through sketches, such as gradient effects and advanced drawing abilities. The entire process of sketching took about 90 minutes. We asked each group to introduce their chart content and embellishment suggestions after they were finished. Following the workshop, participants used a 7-point Likert scale to rate the utility and usability of our design area. Following each group finished sketching, we did semi-interviews with the participants, and some interviews were conducted after the workshop. We wanted to know how participants felt about utilizing ornamentation for data charts, specifically how difficult it was for them to create graphics without the design space and whether our design space helped them better their graphics.

\subsection{Result Analysis}
After the workshop, we collected participants' ratings on the usefulness, usability, and the satisfaction of the outcome. We then conducted in-depth interviews with each participant to understand their ratings and the reasons behind the scores. The collected feedback and analysis are summarized in this section. We use an odd number to indicate an amateur participant and an even number for an expert participant.

\subsubsection{Usefulness of design space.}
Statistics on the method cards show that the participants applied more abundant techniques on each picture after training, and the average types of techniques used in each drawing increased from 5.3 to 6.65. At the same time, the total times of techniques used after training increased (from 106 in 23 types to 130 in 31 types), while the average times decreased (from 4.608 to 4.194). We found the techniques that experts did not seem to increase after the training while amateurs tend to apply more techniques. For example, two visualizations before and after applying the design space are shown in Fig. \ref{fig:workshop}. During the interviews, all the participants asserted that these changes were inspired by the design space and expressed their appreciation of the systematic guidance in their embellishment creations. 
P3 mentioned that ``with the design space, I finally had some clues how to systematically designing an appropriate embellishment.'' P6 also commented, ``I had never seriously considered the communication goals and techniques of embellishments, but this design space give me a chance to better organize design ideas so that I could construct the visualization more effectively.''

Specifically for each method card, the most commonly used techniques after the introduction of the design space are: 1) adding background mask (increase 4 times, 133\%), 2) overlaying contours with different solid colors (increase 3 times, 300\%), and 3) applying two-color systems (increase 3 times, 75\%). There were also eight emerging techniques, such as combining background mask with translucent element, applying the semi-flat style, and superimposing elements on an image. 
The total number of applied technique types increased from 12 to 15 for exploring data, from 11 to 14 for integrating context, and from 0 to 2 for enhancing aesthetics.
However, the most popular technique cards remain the same after introducing the design space: ``fill solid color'', ``overlay contour'', ``apply multiple colors'', and ``add cut-out images''. This indicates that the design space does not completely change the design habits, but provide some suggestions. Therefore, H1 is partially accepted since the diversity indeed increased, especially for amateurs, while experts have their habits or tastes for applying various techniques. P4 claimed that \textit{``the biggest change after using the design space is to focus more on the data itself. In the beginning, I just wanted to emphasize a scene. Later on, I could apply the technique more accurately with the information I want to present.''} P11 mentioned its usefulness for amateur designers, \textit{``I was rather reckless at first when designing embellishments, and the design space reminded me of many potential improvements.''}

When we consider the professional backgrounds of the participants, we found that the richness of embellishment techniques applied indeed increased but the degree of change differed. After the training of using VizBelle, the average number of techniques used by the participants with technical backgrounds increased from 4.3 to 6.3, while this number for the participants with design backgrounds changed from 6.3 to 6.7. The results present an increase in the number of techniques used by all the participants. The richness of the embellishment techniques applied by the users with technical backgrounds was raised significantly, which is much closer to the richness in designers' works.

In the evaluation of the usefulness of the design space, the ratings mostly fall in a similar range (5-6 points). Participants agreed that the design space provides a holistic summary of the communication goals \inlinegraphics{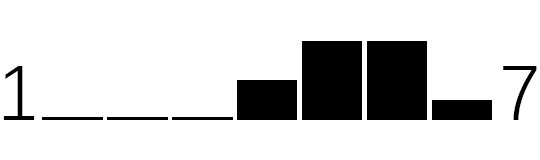}(M=5, SD=0.9), embellishment objects \inlinegraphics{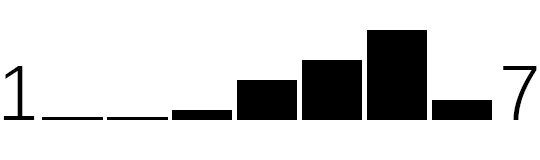}(M=5, SD=1.04) and embellishment techniques \inlinegraphics{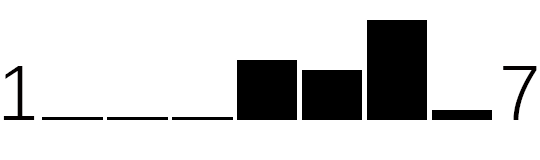} (M=5, SD=0.94) in a clear and logical way \inlinegraphics{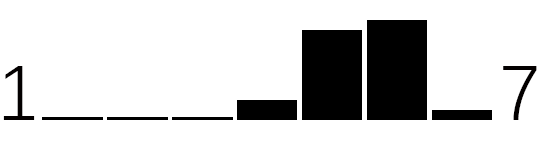}(M=5, SD=0.74). They confirmed its assistance in fast conceiving\inlinegraphics{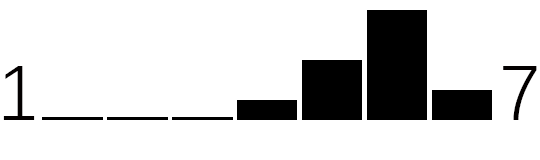} (M=6, SD=0.84), improves efficiency\inlinegraphics{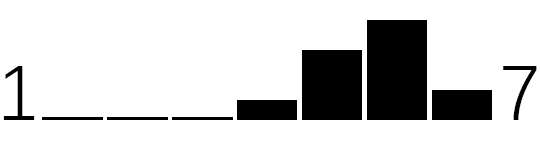} (M=6, SD=0.85) and enables quick determination\inlinegraphics{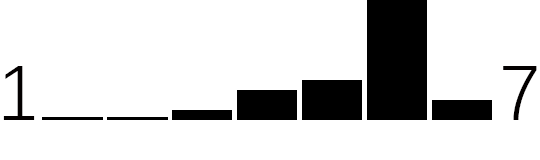}(M=6, SD=1.01). P14 commented that \textit{``the use process of the design space is clear and reasonable''}. However, P19 argued that \textit{``I don't know where to start reading and how to embellish in steps.''} In the interview, P19 explained that he gave the logic and the efficiency dimensions low scores as he hurried to complete the visualization due to the limited time, failing in paying enough attention at the tutorial. 

\subsubsection{Easy to use.}
Since we applied a 7-point Likert scale where 7 is ``strongly agree'' and 1 is ``strongly disagree'', a score of 4 is used as the threshold to evaluate the usability of the design space. A score higher than 4 indicates that design space is regarded as easy to use. The results show that 68\%, 77\%, and 68\% of the participants agreed on the easy-to-use design space where H3 holds. They all praised the use of examples and the easy-to-understand website to facilitate quick search and explanation while they mentioned the difficulty in the actual application of various techniques. To investigate further, we compared the ratings from the experts and the amateur participants to check the instructive tendency of our design space. Amateurs gave higher scores for ``instructiveness for practice'' \inlinegraphics{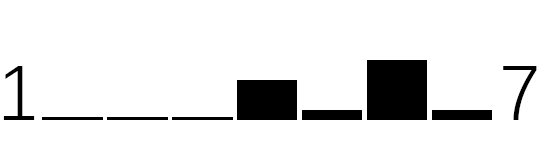} (M=5.3, SD=1.07) compared to experts \inlinegraphics{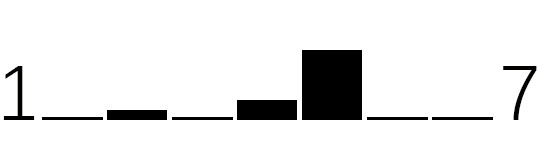} (M=4.5, SD=0.92). This may be because experts have established their own way of working with data embellishments. Surprisingly, amateurs rated higher than experts in all the three dimensions for usability. This confirmed that our design space is friendly for amateurs, but might not contribute enough to the experts' design process.
In general, all the participants were satisfied with the overall effect \inlinegraphics{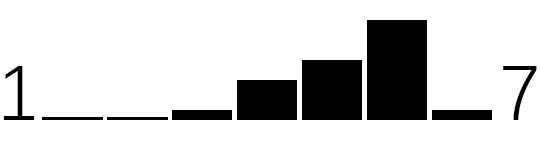} (M=5.27, SD=0.98), willing to continue using this tool \inlinegraphics{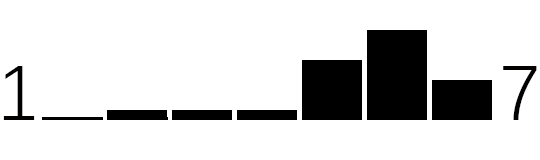} (M=5.5, SD=1.26). They agreed that the design space fulfilled most of their requirements \inlinegraphics{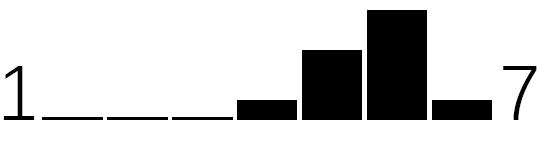} (M=5.59, SD=0.8) and were happy to recommend it to others in the future \inlinegraphics{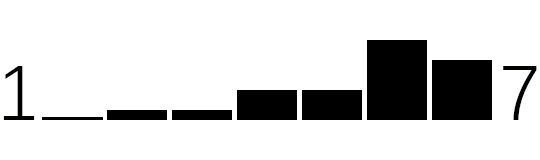} (M=5.55, SD=1.4).

\subsubsection{User satisfaction of outcome}
This section mainly compares user satisfaction of the design outcomes before and after using the proposed design space with the Wilcoxon signed-rank tests.
As shown in Table. \ref{tab:difference}, the user satisfaction of the design outcomes has been significantly improved in all the dimensions we evaluated. In the evaluation of self-performance, apart from the insight exploration, user satisfaction has also improved significantly. H2 is accepted. 
Through further comparison on the evaluations of the design works by amateurs and experts before and after training, we found that the average score of amateurs has increased by 23\%, and the average score of experts has increased by 27\%. Thus, H4 is rejected. 

\begin{table*}[!tb]
\caption{Scores of user satisfaction before and after using the design space.}
\resizebox{\textwidth}{!}{%
\begin{tabular}{@{}lc|ccccccccccccccccl@{}}
\toprule
 & \multirow{3}{*}{User Satisfaction} & \multicolumn{8}{c}{Self-performance Satisfaction} & \multicolumn{8}{c}{Design Outcome Satisfaction} &  \\ \cmidrule(l){3-19} 
 &  & \multicolumn{2}{c|}{Data Insights} & \multicolumn{2}{c|}{Creative Thinking} & \multicolumn{2}{c|}{Object Selection} & \multicolumn{2}{c|}{Overall Experience} & \multicolumn{2}{c|}{Overall Effect} & \multicolumn{2}{c|}{Diversity} & \multicolumn{2}{c|}{Creativity} & \multicolumn{2}{c}{Efficiency} &  \\ \cmidrule(l){3-19} 
 &  & before & \multicolumn{1}{c|}{after} & before & \multicolumn{1}{c|}{after} & before & \multicolumn{1}{c|}{after} & before & \multicolumn{1}{c|}{after} & before & \multicolumn{1}{c|}{after} & before & \multicolumn{1}{c|}{after} & before & \multicolumn{1}{c|}{after} & before & after &  \\ \midrule
 & Median & 5 & \multicolumn{1}{c|}{5} & 4 & \multicolumn{1}{c|}{6} & 5 & \multicolumn{1}{c|}{5} & 5 & \multicolumn{1}{c|}{6} & 5 & \multicolumn{1}{c|}{6} & 4 & \multicolumn{1}{c|}{6} & 4 & \multicolumn{1}{c|}{6} & 4.5 & 6 &  \\
 & Variance & 1.134 & \multicolumn{1}{c|}{1.275} & 1.206 & \multicolumn{1}{c|}{0.736} & 1.751 & \multicolumn{1}{c|}{1.180} & 2.355 & \multicolumn{1}{c|}{1.015} & 1.784 & \multicolumn{1}{c|}{1.022} & 1.950 & \multicolumn{1}{c|}{0.799} & 2.190 & \multicolumn{1}{c|}{1.004} & 2.087 & 1.110 &  \\
 & Mean & 4.909 & \multicolumn{1}{c|}{5.318} & 4.409 & \multicolumn{1}{c|}{5.545} & 4.682 & \multicolumn{1}{c|}{5.318} & 4.545 & \multicolumn{1}{c|}{5.409} & 4.545 & \multicolumn{1}{c|}{5.455} & 4.045 & \multicolumn{1}{c|}{5.682} & 4.000 & \multicolumn{1}{c|}{5.636} & 4.091 & 5.409 &  \\ \midrule
 
 & P-value & \multicolumn{2}{c|}{0.164} & \multicolumn{2}{c|}{0.000557215} & \multicolumn{2}{c|}{0.014305878} & \multicolumn{2}{c|}{0.011783220} & \multicolumn{2}{c|}{0.005099317} & \multicolumn{2}{c|}{0.000252201} & \multicolumn{2}{c|}{0.000090471} & \multicolumn{2}{c}{0.000795053} &  \\ \bottomrule
\end{tabular}%
}
\label{tab:difference}
\end{table*}


\textit{\textbf{Convey data insights.}}
The improvement in insight exploration of data is not obvious. One reason is that the workshop has provided clear data facts and baseline charts. Another reason is that many users do not regard ``presenting'' data facts as a part of data exploration. Users generally believe that it is not difficult to understand the data facts given in the workshop, so there is no obvious improvement. However, P7 and P14 pointed out that the design space can inspire themselves to \textit{``adjust data encoding''} or \textit{``highlight targets in complex information''}.

\textit{\textbf{Engage readers in the context.}}
The design space provides significant help in improving the figurative design. For example, P2 shared his experience of drawing a specific visual embellishment, \textit{``in order to make the visualization more attractive, I changed the original histogram to a legend that is closer to the shape of the gold medal.''}. P11 pointed out, \textit{``the design space gave me hints about inserting images to the background, and even in the form of a mask, so I changed the points on the line graph to make them look like a movie-play button.''} 

\textit{\textbf{Enhance aesthetics.}}
All the participants agreed that design space helps improve the aesthetic level. Five users emphasized this aspect, stating that this part of the design space assisted them in {``improving the ornamental and entertaining effects of the chart''}. Since the workshop was conducted offline, almost all the participants applied the hand-drawn comic style while few participants attempted other styles with limited tools. P18 applied the MBE style, P1 and P12 used similar color schemes to indicate gradient color, P2 described his plan of using a semi-flat aesthetic effect in the text design. Several participants reported that \textit{``some techniques are not easy to implement by hand so I did not apply them during the workshop, but would like to try out later at work''}, such as applying gradient effects, adding photos, conducting multi-layer display, and so on. 

\section{Discussion}
In this section, we discuss the common usage patterns of our design space in the workshop, the generalizability issue, the implications for designing embellishment tools, the limitations and future work.

\subsection{Common Usage Patterns.}
Based on the interview results and the design outcomes, we identified 6 design space usage patterns as follows.

\textit{\textbf{Standard} (goal-object-technique)}: 25\% of the participants applied the standard scheme introduced in the workshop. The ratio of technical background versus design background, and amateurs versus professionals are both 4:1. It shows that the standard strategy is the most popular among amateur practitioners with technical backgrounds. 

\textit{\textbf{Half-divergent} (goal-technique)}: 15\% of the participants adopted the half-divergent scheme, all had design backgrounds. Experts among them believed that \textit{``I do not need to specify the embellishment objects to draw my works''}, while amateurs put forward that \textit{``the column of embellishment objects is a bit difficult to understand at the beginning, but it helped later on when I check potential targets for applying various techniques''}. 

\textit{\textbf{Bidirectional} (goal-technique-object)}: 10\% of the participants adopted the bidirectional scheme, all were amateur designers. They were not confident with their ideas and constantly searched for various techniques from the examples. P9 mentioned that \textit{``the most difficult thing for me is the ideation, that is, how to transform the simple data chart into a more attractive graphical form. I do not have much experience in exploring data insights, and I rely a lot on the example cases from the website to finish my design.''}

\textit{\textbf{Inspirational} (technique-object)}: 25\% of the participants used inspirational scheme, of which 80\% had technical backgrounds, and 20\% had the design backgrounds. In terms of practical experience, experts accounted for 60\% and amateurs accounted for 40\%. Most users adopting this scheme stated that the standard scheme did not completely match their own workflow. However, they could still get inspirations from the design space and the cases provided on the website. 

\textit{\textbf{Contrary} (technique-goal)}: 15\% of the participants used the contrary strategy, and all were designers. P13 claimed that \textit{``my designs were completely based on random ideas and intentions before the workshop training, and the design space helped me organize my thoughts.''} P16 pointed out that \textit{``going through all the techniques helps me broaden my design choices. The design space seems comprehensive, but sometimes it is a bit confusing for me that the same technique serves different goals. I need to spare extra time to figure out how differently it is used."}

\textit{\textbf{Occasional} (random usage as a reference)}: 10\% of the participants did not actually follow the design space, who were all expert practitioners. In most cases, they still relied on their own experience and used it as a reference to complement their designs when they got stuck. 

\subsection{The Generalizability of the Design Space.}
Given that infographics are considered a static, widely-used form of data visualization, they can also inspire other visualization genres\cite{segel2010narrative}. For example, the use of animation in data videos is very similar to that of visual embellishment in infographics, both of which can be used to reveal transformative data facts, convey emotions, facilitate communication and enhance engagement\cite{shi2021communicating}. Therefore, animation in data videos can be regarded as a dynamic extension of visual embellishments. Moreover, in magazines, slides, and sketch notes, infographics are the main carrier of quantitative analysis, and visual embellishments are important tools for expressing information and guiding viewers' attention. Our results can be easily generalized to the magazine-style for the similarities between infographics and magazine pages in terms of layout and content. In slide decks that combine plotted charts and captions, VizBelle can provide a basis for creating templates that are systematic and customizable. It can also be linked with a recommendation system to form an advanced design tool. In sketch notes, the strategies to capture ideas, emphasize key concepts, create a narrative, improve aesthetics and visual identity, and deal with constraints of live drawing are identical to those presented by our workshop participants. 
In short, communicating data requires tailoring messages to those receiving it; thus visual embellishments that provide concise and clearly structured messages in delicate ways can be quite inspiring. Therefore, VizBelle can also be taken as a framework of practice in cognitive science \cite{KosslynImageBrainResolution1996}. It puts forward a preliminary method that can minimize cognitive bias as well as cognitive efficiency through visual cues and visual metaphors. Applications in education and communications studies may benefit from our results \cite{SpectorHandbookResearchEducational2014}.

\subsection{Implications for Designing Embellishment Tools.}
Existing tools can help create embellishments by adding pictures or pictograms to standard visualizations and providing suggestions and recommendations at the ideation stage. For example, MetaMap~\cite{kang2021metamap} provides example-based exploration in semantics, color, and shape to generate creative ideas.
However, as mentioned in the expert interviews, the process of applying embellishments to visualization is non-linear. Nevertheless, direct application of our design space to the design and implementation of embellishment creation tools remains to be a challenging problem. We summarized several usage patterns from the workshop, which can provide some ideas to design recommendation algorithms for different types of users. Based on our design space design and usage patterns, future opportunities include developing automatic or semi-automatic tools for different types of users that employ the recommendation algorithm to assist in the generation of appropriate and engaging visual embellishments. 
Moreover, since the correct use of various embellishment techniques require long-term practice, it would be valuable to provide tools that inform readers of any misuse in the embellishment which can misguide them. Such tools can also be used in teaching visualization and to help train future designers. We hope that this design space could be used a starting guideline to embed state-of-art techniques and tools to help facilitate the whole ideation and creation process.

\subsection{Limitations and Future Work.}
Although we received much positive feedback, this work still has several limitations.
First, our collected corpus is quite varied and contains up-to-date infographics. However, the current sources are mostly from data journalism, which may exclude some complex and artistic embellishments. We can further expand the corpus by collecting more cases and incorporating other design elements from various resources. For example, collecting award-winning works in data visualization over the years and selected design works with data insights from design books may be useful. 
Second, amateurs and experts have various purposes and demand different guidance and assistance. On the one hand, many amateurs did not know the general principles for visual encoding. Accordingly, many amateurs encountered difficulties at the conception stage. The optimal goal is to design innovative and interesting visualizations with both reasonable visual encoding and attractive visual embellishments. Therefore, how to combine techniques in visual encoding and data embellishment is an important future research direction. On the other hand, the benefits of the design space are limited for expert designers. In the future, we plan to provide more advanced functions, such as layout techniques based on the arrangement logic of multiple charts, or adding interactive or dynamic embellishing methods. 
Third, a systematic evaluation on the final output is lacking. We still cannot know whether the final design output achieved the authors' communication goals because the current outcome satisfaction scores were marked by the creators. Thus, another user study with a different group of participants to evaluate the workshop design outcomes can further verify the effectiveness of the design space.

\section{Conclusion}

In this paper, we presented VizBelle, a design space to help designers create embellishments in data visualization. We derived three major communication goals, identified 21 strategies and corresponding techniques to fulfill different goals from the analysis of 361 infographics. We conducted a workshop to observe the usage patterns and the usefulness of VizBelle. The results proved that the design space could support amateur and professional designers in generating more insightful and diverse ideas for decorating data visualizations. We hope our work could provide new perspectives and more inspirations in the ideation and creation process to facilitate various design activities.




\bibliographystyle{abbrv-doi}

\bibliography{idvxlab}

\begin{thebibliography}{10}

\bibitem{AdornoCultureIndustrySelected2005}
T.~W. Adorno and J.~M. Bernstein.
\newblock {\em The {{Culture Industry}}: {{Selected Essays}} on {{Mass
  Culture}}.}
\newblock {Taylor and Francis}, {Florence}, 2005.

\bibitem{ajani2021declutter}
K.~Ajani, E.~Lee, C.~Xiong, C.~N. Knaflic, W.~Kemper, and S.~Franconeri.
\newblock Declutter and focus: Empirically evaluating design guidelines for
  effective data communication.
\newblock {\em IEEE Transactions on Visualization and Computer Graphics}, 2021.

\bibitem{andry2021interpreting}
T.~Andry, C.~Hurter, F.~Lambotte, P.~Fastrez, and A.~Telea.
\newblock Interpreting the effect of embellishment on chart visualizations.
\newblock In {\em Proceedings of the 2021 CHI Conference on Human Factors in
  Computing Systems}, pp. 1--15, 2021.

\bibitem{AumontImage1997}
J.~Aumont.
\newblock {\em The {{Image}}}.
\newblock {British Film Institute}, 1997.

\bibitem{avraamidou2009role}
L.~Avraamidou and J.~Osborne.
\newblock The role of narrative in communicating science.
\newblock {\em International Journal of Science Education}, 31(12):1683--1707,
  2009.

\bibitem{BarnardGraphicDesignCommunication2013}
M.~Barnard.
\newblock {\em Graphic {{Design}} as {{Communication}}}.
\newblock {Routledge}, zeroth ed., July 2013. doi: {{%
10\hspace{.1pt}\discretionary{.}{%
}{.}\hspace{.4pt}4324\discretionary{/}{%
}{/}9781315015385}}


\bibitem{NAPACardsNarrative}
L.~Bartram, J.~Boy, P.~Ciuccarelli, S.~Drucker, Y.~Engelhardt, U.~Koeppen,
  M.~Stefaner, B.~Tversky, and J.~Wood.
\newblock {{NAPA Cards}} \_ {{Narrative Patterns}} for {{Data Stories}}.
\newblock http://napa-cards.net/\#info.
\newblock Accessed: March 31, 2022.

\bibitem{bateman2014text}
J.~A. Bateman.
\newblock {\em Text and image: A critical introduction to the visual/verbal
  divide}.
\newblock Routledge, 2014.

\bibitem{bateman2010useful}
S.~Bateman, R.~L. Mandryk, C.~Gutwin, A.~Genest, D.~McDine, and C.~Brooks.
\newblock Useful junk? the effects of visual embellishment on comprehension and
  memorability of charts.
\newblock In {\em Proceedings of the SIGCHI conference on human factors in
  computing systems}, pp. 2573--2582, 2010.

\bibitem{BertinSemiologyGraphics1983}
J.~Bertin.
\newblock {\em Semiology of {{Graphics}}}.
\newblock {University of Wisconsin Press}, 1983.

\bibitem{best2008taxonomy}
R.~Best, R.~Floyd, and D.~McNamara.
\newblock Taxonomy of educational objectives. handbook 1: Cognitive domain.
\newblock {\em Reading Psychology}, 29(2):137--164, 2008.

\bibitem{BorgoEmpiricalStudyUsing2012}
R.~Borgo, A.~{Abdul-Rahman}, F.~Mohamed, P.~W. Grant, I.~Reppa, L.~Floridi, and
  {Min Chen}.
\newblock An {{Empirical Study}} on {{Using Visual Embellishments}} in
  {{Visualization}}.
\newblock {\em IEEE Transactions on Visualization and Computer Graphics},
  18(12):2759--2768, Dec. 2012. doi: {{%
10\hspace{.1pt}\discretionary{.}{%
}{.}\hspace{.4pt}1109\discretionary{/}{%
}{/}TVCG\hspace{.1pt}\discretionary{.}{%
}{.}\hspace{.4pt}2012\hspace{.1pt}\discretionary{.}{%
}{.}\hspace{.4pt}197}}


\bibitem{borkin2015beyond}
M.~A. Borkin, Z.~Bylinskii, N.~W. Kim, C.~M. Bainbridge, C.~S. Yeh, D.~Borkin,
  H.~Pfister, and A.~Oliva.
\newblock Beyond memorability: Visualization recognition and recall.
\newblock {\em IEEE transactions on visualization and computer graphics},
  22(1):519--528, 2015.

\bibitem{BrunerActualmindspossible1986}
J.~S. Bruner.
\newblock {\em Actual Minds, Possible Worlds}.
\newblock {Harvard University Press}, {Cambridge, Mass}, 1986.

\bibitem{bryan2016temporal}
C.~Bryan, K.-L. Ma, and J.~Woodring.
\newblock Temporal summary images: An approach to narrative visualization via
  interactive annotation generation and placement.
\newblock {\em IEEE transactions on visualization and computer graphics},
  23(1):511--520, 2016.

\bibitem{burgio2017infographics}
V.~Burgio and M.~Moretti.
\newblock Infographics as images: Meaningfulness beyond information.
\newblock In {\em Multidisciplinary Digital Publishing Institute Proceedings},
  vol.~1, p. 891, 2017.

\bibitem{burke2009isotype}
C.~Burke.
\newblock Isotype representing social facts pictorially.
\newblock {\em Information Design Journal}, 17(3):211--223, 2009.

\bibitem{burns2021designing}
A.~Burns, C.~Xiong, S.~Franconeri, A.~Cairo, and N.~Mahyar.
\newblock Designing with pictographs: Envision topics without sacrificing
  understanding.
\newblock {\em IEEE transactions on visualization and computer graphics}, 2021.

\bibitem{BusselleMeasuringNarrativeEngagement2009}
R.~Busselle and H.~Bilandzic.
\newblock Measuring {{Narrative Engagement}}.
\newblock {\em Media Psychology}, 12(4):321--347, Nov. 2009. doi: {{%
10\hspace{.1pt}\discretionary{.}{%
}{.}\hspace{.4pt}1080\discretionary{/}{%
}{/}15213260903287259}}


\bibitem{byrne2019figurative}
L.~Byrne, D.~Angus, and J.~Wiles.
\newblock Figurative frames: A critical vocabulary for images in information
  visualization.
\newblock {\em Information Visualization}, 18(1):45--67, 2019.

\bibitem{cairo2012functional}
A.~Cairo.
\newblock {\em The Functional Art: An introduction to information graphics and
  visualization}.
\newblock New Riders, 2012.

\bibitem{canham2010effects}
M.~Canham and M.~Hegarty.
\newblock Effects of knowledge and display design on comprehension of complex
  graphics.
\newblock {\em Learning and instruction}, 20(2):155--166, 2010.

\bibitem{cleveland1993visualizing}
W.~S. Cleveland.
\newblock {\em Visualizing data}.
\newblock Hobart press, 1993.

\bibitem{cruz2015wrongfully}
P.~Cruz.
\newblock Wrongfully right: applications of semantic figurative metaphors in
  information visualization.
\newblock {\em IEEE VIS Arts Program (VISAP)}, pp. 14--21, 2015.

\bibitem{CruzSemanticfigurativemetaphors}
P.~Cruz.
\newblock {\em Semantic Figurative Metaphors in Information Visualization}.
\newblock PhD thesis, Universidade de Coimbra (Portugal), 2016.

\bibitem{cui2019text}
W.~Cui, X.~Zhang, Y.~Wang, H.~Huang, B.~Chen, L.~Fang, H.~Zhang, J.-G. Lou, and
  D.~Zhang.
\newblock Text-to-viz: Automatic generation of infographics from
  proportion-related natural language statements.
\newblock {\em IEEE transactions on visualization and computer graphics},
  26(1):906--916, 2019.

\bibitem{Design000IconsSymbols2006}
B.~Design.
\newblock {\em 1,000 {{Icons}}, {{Symbols}}, and {{Pictograms}}: {{Visual
  Communications}} for {{Every Language}}}.
\newblock {Rockport Publishers}, Sept. 2006.

\bibitem{DuarteDatastoryexplain2019}
N.~Duarte.
\newblock {\em Data Story: Explain Data and Inspire Action through Story}.
\newblock {Ideapress Publishing}, {Oakton, Virginia}, 2019.

\bibitem{engebretsen2020data}
M.~Engebretsen and H.~Kennedy.
\newblock {\em Data visualization in society}.
\newblock Amsterdam University Press., 2020.

\bibitem{FewChartjunkDebateClose2011}
S.~Few.
\newblock The {{Chartjunk Debate}} \textendash{} {{A Close Examination}} of
  {{Recent Findings}}.
\newblock {\em Perceptual Edge}, p.~11, 2011.

\bibitem{FewInformationVisualizationResearch2015a}
S.~Few.
\newblock Information {{Visualization Research}} as {{Pseudo}}-{{Science}}.
\newblock {\em Perceptual Edge}, p.~9, 2015.

\bibitem{figueiras2018review}
A.~Figueiras.
\newblock A review of visualization assessment in terms of user performance and
  experience.
\newblock In {\em 2018 22nd International Conference Information Visualisation
  (IV)}, pp. 145--152. IEEE, 2018.

\bibitem{haroz2015isotype}
S.~Haroz, R.~Kosara, and S.~L. Franconeri.
\newblock Isotype visualization: Working memory, performance, and engagement
  with pictographs.
\newblock In {\em Proceedings of the 33rd annual ACM conference on human
  factors in computing systems}, pp. 1191--1200, 2015.

\bibitem{holmes2000pictograms}
N.~Holmes.
\newblock Pictograms: A view from the drawing board or, what i have learned
  from otto neurath and gerd arntz (and jazz).
\newblock {\em Information design journal}, 10(2):133--143, 2000.

\bibitem{hou2020rhetorical}
S.~Hou, S.~Zhang, and C.~Fei.
\newblock Rhetorical structure theory: A comprehensive review of theory,
  parsing methods and applications.
\newblock {\em Expert Systems with Applications}, 157:113421, 2020.

\bibitem{hullman2011benefitting}
J.~Hullman, E.~Adar, and P.~Shah.
\newblock Benefitting infovis with visual difficulties.
\newblock {\em IEEE Transactions on Visualization and Computer Graphics},
  17(12):2213--2222, 2011.

\bibitem{hullman2011visualization}
J.~Hullman and N.~Diakopoulos.
\newblock Visualization rhetoric: Framing effects in narrative visualization.
\newblock {\em IEEE transactions on visualization and computer graphics},
  17(12):2231--2240, 2011.

\bibitem{MethodCards}
IDEO.
\newblock Method {{Cards}}.
\newblock https://www.ideo.com/post/method-cards.
\newblock Accessed: March 31, 2022.

\bibitem{kandogan2012just}
E.~Kandogan.
\newblock Just-in-time annotation of clusters, outliers, and trends in
  point-based data visualizations.
\newblock In {\em 2012 IEEE Conference on Visual Analytics Science and
  Technology (VAST)}, pp. 73--82. IEEE, 2012.

\bibitem{kang2021metamap}
Y.~Kang, Z.~Sun, S.~Wang, Z.~Huang, Z.~Wu, and X.~Ma.
\newblock Metamap: Supporting visual metaphor ideation through
  multi-dimensional example-based exploration.
\newblock In {\em Proceedings of the 2021 CHI Conference on Human Factors in
  Computing Systems}, pp. 1--15, 2021.

\bibitem{kelliher2012tell}
A.~Kelliher and M.~Slaney.
\newblock Tell me a story.
\newblock {\em IEEE multimedia}, 19(1):4--4, 2012.

\bibitem{KosslynImageBrainResolution1996}
S.~M. Kosslyn.
\newblock {\em Image and {{Brain}}: {{The Resolution}} of the {{Imagery
  Debate}}}.
\newblock {MIT Press}, 1996.

\bibitem{KrumCoolinfographicseffective2014}
R.~Krum.
\newblock {\em Cool Infographics: Effective Communication with Data
  Visualization and Design}.
\newblock {Wiley}, {Indianapolis, IN}, 2014.

\bibitem{LankowInfographicsPowerVisual2012}
J.~Lankow, J.~Ritchie, and R.~Crooks.
\newblock {\em Infographics: {{The Power}} of {{Visual Storytelling}}}.
\newblock {John Wiley \& Sons}, Sept. 2012.

\bibitem{latif2021deeper}
S.~Latif, S.~Chen, and F.~Beck.
\newblock A deeper understanding of visualization-text interplay in geographic
  data-driven stories.
\newblock In {\em Computer Graphics Forum}, vol.~40, pp. 311--322. Wiley Online
  Library, 2021.

\bibitem{lau2007towards}
A.~Lau and A.~V. Moere.
\newblock Towards a model of information aesthetics in information
  visualization.
\newblock In {\em 2007 11th International Conference Information Visualization
  (IV'07)}, pp. 87--92. IEEE, 2007.

\bibitem{li2014chart}
H.~Li and N.~Moacdieh.
\newblock Is “chart junk” useful? an extended examination of visual
  embellishment.
\newblock In {\em Proceedings of the Human Factors and Ergonomics Society
  Annual Meeting}, vol.~58, pp. 1516--1520. Sage Publications Sage CA: Los
  Angeles, CA, 2014.

\bibitem{lu2020exploring}
M.~Lu, C.~Wang, J.~Lanir, N.~Zhao, H.~Pfister, D.~Cohen-Or, and H.~Huang.
\newblock Exploring visual information flows in infographics.
\newblock In {\em Proceedings of the 2020 CHI conference on human factors in
  computing systems}, pp. 1--12, 2020.

\bibitem{moere2012evaluating}
A.~V. Moere, M.~Tomitsch, C.~Wimmer, B.~Christoph, and T.~Grechenig.
\newblock Evaluating the effect of style in information visualization.
\newblock {\em IEEE transactions on visualization and computer graphics},
  18(12):2739--2748, 2012.

\bibitem{NormanEmotionalDesignWhy2004}
D.~A. Norman.
\newblock {\em Emotional {{Design}}: {{Why We Love}} (or {{Hate}}) {{Everyday
  Things}}}.
\newblock {Basic Books}, 2004.

\bibitem{ojo2018patterns}
A.~Ojo and B.~Heravi.
\newblock Patterns in award winning data storytelling: Story types, enabling
  tools and competences.
\newblock {\em Digital journalism}, 6(6):693--718, 2018.

\bibitem{passonneau2014benefits}
R.~J. Passonneau and B.~Carpenter.
\newblock The benefits of a model of annotation.
\newblock {\em Transactions of the Association for Computational Linguistics},
  2:311--326, 2014.

\bibitem{ren2017chartaccent}
D.~Ren, M.~Brehmer, B.~Lee, T.~H{\"o}llerer, and E.~K. Choe.
\newblock Chartaccent: Annotation for data-driven storytelling.
\newblock In {\em 2017 IEEE Pacific Visualization Symposium (PacificVis)}, pp.
  230--239. IEEE, 2017.

\bibitem{RicheDatadrivenstorytelling2018}
N.~H. Riche, C.~Hurter, N.~Diakopoulos, and S.~Carpendale, eds.
\newblock {\em Data-Driven Storytelling}.
\newblock A {{K Peters Visualization Series}}. {CRC Press/Taylor \& Francis
  Group}, {Boca Raton, Florida}, 2018.

\bibitem{romat2020dear}
H.~Romat, N.~Henry~Riche, C.~Hurter, S.~Drucker, F.~Amini, and K.~Hinckley.
\newblock Dear pictograph: Investigating the role of personalization and
  immersion for consuming and enjoying visualizations.
\newblock In {\em Proceedings of the 2020 CHI Conference on Human Factors in
  Computing Systems}, pp. 1--13, 2020.

\bibitem{RosenbergAuToBIToolAutomatic}
A.~Rosenberg.
\newblock {{AuToBI}} --- {{A Tool}} for {{Automatic ToBI Annotation}}.
\newblock In {\em Eleventh Annual Conference of the International Speech
  Communication Association}, p.~4, 2010.

\bibitem{saket2016beyond}
B.~Saket, A.~Endert, and J.~Stasko.
\newblock Beyond usability and performance: A review of user experience-focused
  evaluations in visualization.
\newblock In {\em Proceedings of the Sixth Workshop on Beyond Time and Errors
  on Novel Evaluation Methods for Visualization}, pp. 133--142, 2016.

\bibitem{segel2010narrative}
E.~Segel and J.~Heer.
\newblock Narrative visualization: Telling stories with data.
\newblock {\em IEEE transactions on visualization and computer graphics},
  16(6):1139--1148, 2010.

\bibitem{shi2020calliope}
D.~Shi, X.~Xu, F.~Sun, Y.~Shi, and N.~Cao.
\newblock Calliope: Automatic visual data story generation from a spreadsheet.
\newblock {\em IEEE Transactions on Visualization and Computer Graphics},
  27(2):453--463, 2020.

\bibitem{shi2021communicating}
Y.~Shi, X.~Lan, J.~Li, Z.~Li, and N.~Cao.
\newblock Communicating with motion: A design space for animated visual
  narratives in data videos.
\newblock In {\em Proceedings of the 2021 CHI Conference on Human Factors in
  Computing Systems}, pp. 1--13, 2021.

\bibitem{skau2015evaluation}
D.~Skau, L.~Harrison, and R.~Kosara.
\newblock An evaluation of the impact of visual embellishments in bar charts.
\newblock In {\em Computer Graphics Forum}, vol.~34, pp. 221--230. Wiley Online
  Library, 2015.

\bibitem{SpectorHandbookResearchEducational2014}
J.~M. Spector, M.~D. Merrill, J.~Elen, and M.~J. Bishop, eds.
\newblock {\em Handbook of {{Research}} on {{Educational Communications}} and
  {{Technology}}}.
\newblock {Springer New York}, {New York, NY}, 2014. doi: {{%
10\hspace{.1pt}\discretionary{.}{%
}{.}\hspace{.4pt}1007\discretionary{/}{%
}{/}978\discretionary{%
}{-}{-}1\discretionary{%
}{-}{-}4614\discretionary{%
}{-}{-}3185\discretionary{%
}{-}{-}5}}


\bibitem{thibodeau2019role}
P.~H. Thibodeau, T.~Matlock, and S.~J. Flusberg.
\newblock The role of metaphor in communication and thought.
\newblock {\em Language and Linguistics Compass}, 13(5):e12327, 2019.

\bibitem{TufteEnvisioningInformation1990}
E.~R. Tufte.
\newblock {\em Envisioning {{Information}}}.
\newblock {Graphics Press}, 1990.

\bibitem{TufteVisualDisplayQuantitative2001}
E.~R. Tufte.
\newblock {\em The {{Visual Display}} of {{Quantitative Information}}}.
\newblock {Graphics Press}, 2001.

\bibitem{wang2019datashot}
Y.~Wang, Z.~Sun, H.~Zhang, W.~Cui, K.~Xu, X.~Ma, and D.~Zhang.
\newblock Datashot: Automatic generation of fact sheets from tabular data.
\newblock {\em IEEE transactions on visualization and computer graphics},
  26(1):895--905, 2019.

\bibitem{xiong2019curse}
C.~Xiong, L.~Van~Weelden, and S.~Franconeri.
\newblock The curse of knowledge in visual data communication.
\newblock {\em IEEE transactions on visualization and computer graphics},
  26(10):3051--3062, 2019.

\bibitem{YiUnderstandingCharacterizingInsights}
J.~S. Yi, Y.-a. Kang, J.~T. Stasko, and J.~A. Jacko.
\newblock Understanding and {{Characterizing Insights}}: {{How Do People Gain
  Insights Using Information Visualization}}?
\newblock In {\em Proceedings of the 2008 Workshop on BEyond time and errors:
  novel evaLuation methods for Information Visualization}, pp. 1--6, 2008.

\bibitem{zacks1998reading}
J.~Zacks, E.~Levy, B.~Tversky, and D.~J. Schiano.
\newblock Reading bar graphs: Effects of extraneous depth cues and graphical
  context.
\newblock {\em Journal of experimental psychology: Applied}, 4(2):119, 1998.

\end{thebibliography}
\end{document}